\documentclass[11pt]{article}
\newdimen\trimheight \trimheight9truein
\newdimen\trimwidth \trimwidth6truein
\newdimen\typeheight \typeheight7.35in

\begin{document}

\title{BPS M-Brane Geometries.}
\author{Tasneem Zehra Husain\footnote{email: tasneem@physics.harvard.edu}\\
Jefferson Physical Laboratory, \\
Harvard University, Cambridge, MA 02138} 

\maketitle

\vspace{1cm}

\begin{abstract}
\noindent
In the search for a classification of BPS backgrounds with flux, we look at geometries that arise when M-branes wrap supersymmetric cycles in Calabi-Yau manifolds. We find constraints on the differential forms  in the back-reacted manifolds and discover that the calibration corresponding to the (background generating) M-brane is a co-closed form.\\

\noindent
{\it Contribution to proceedings of the 12th Regional Conference in Mathematical Physics, Islamabad, March 2006.}
\end{abstract}

\vspace{-13cm}
\begin{flushright}
HUTP-06/A0029 \\
hep-th/0607103
\end{flushright}

\thispagestyle{empty}

\newpage

\tableofcontents

\section{Introduction}
Ever since it was found that compactifications of String/M-Theory on special holonomy manifolds preserve supersymmetry, such manifolds have been widely studied.  If Euclidean, they must also be Ricci-flat and according to Berger's classification their holonomy groups are then dictated by their dimension $d$; we have $SU(n)$ holonomy (Calabi-Yau manifolds) in $d=2n$ dimensions,  $Sp(n)$ holonomy (hyper-K\"ahler manifolds) if $d = 4n$, $G_2$ holonomy in $d=7$ and $Spin(7)$ in $d=8$. This neat categorization however, applies only in the absence of flux; as we will now see, supersymmetric backgrounds are no longer quite so simple once space-time contains flux.

\section{D=11 Supergravity}
We will study the geometry of BPS M-branes using 11-d supergravity \cite{sugra}. While admittedly just an approximation to M-theory, supergravity is useful when considering BPS states since these are protected from quantum corrections by supersymmetry and hence guaranteed to survive the transition to strong coupling/M-Theory. 

The bosonic fields of d=11 supergravity are the metric $G_{IJ}$ and a three-form gauge potential A (with associated field strength $F = dA$)  which couples electrically to M2-branes and magnetically to M5-branes. The bosonic action\footnote{ The field content of 11-dimensional supergravity also contains a fermion - the gravitino $\Psi$. However, since we are considering purely bosonic solutions, we are not concerned with how the gravitino appears in the action or its equations of motion.}  of this theory
\begin{equation}
S = \frac{1}{2 {\kappa}^2 } \int d^{11} x \sqrt{-G} \; R - \frac{1}{2} F \wedge * F - \frac{1}{6} A \wedge F \wedge F
\label{action}
\end{equation}
leads to the equation of motion
\begin{equation}
R_{IJ} = \frac{1}{12} F_{IKLM} F_{J}^{\; KLM} - \frac{1}{144} G_{IJ} F^{KLMN} F_{KLMN}
\label{R}
\end{equation}
together with the Bianchi Identity $dF=0$ and equation of motion for the field strength $d*F + {\frac{1}{2}}  F \wedge F = 0$.

In BPS backgrounds the supersymmetric variations of all fields vanish when the variation parameter is a Killing Spinor. Since we are restricting ourselves to purely bosonic backgrounds, we have get the gravitino $\Psi$ to zero, hence the supersymmetric variation of the bosonic fields vanishes trivially. The variation of the gravitino however, is proportional to the bosonic fields and hence not zero a priori.  In order to guarantee supersymmetry, we impose 
\begin{equation}
\delta_{\chi} {\Psi}_I = [ \nabla_I   - {\frac{1}{18}} \Gamma^{ JKL} F_{IJKL} 
+ \frac{1}{144} \Gamma_{I}^{\;\; JKLM} F_{JKLM} ] \chi = 0
\label{psi}
\end{equation}

When $F =0$, the Ricci tensor vanishes and supersymmetry is preserved only if
the background admits covariantly constant spinors. From  $\nabla_I \chi = 0$\footnote{Here, $\nabla_I \chi  = [\partial_I + {\frac{1}{4}} \omega^{ij}_I \Gamma_{ij}] \chi$.}  it follows that 
$ [\nabla_I, \nabla_J] \chi = 0$ and the identitiy $ [\nabla_I, \nabla_J] \chi = \frac{1}{4} R_{IJKL} \Gamma^{KL} \chi$ can then be used to show that a Killing spinor in such a background must be a singlet of the $Spin(1,10)$ subgroup ${\cal H}$ generated by $R_{IJKL} \Gamma^{KL}$. In other words, ${\cal H}$ has to be a special holonomy group. This is the logic that behind the oft-quoted statement that String/M-Theory compactified on special holonomy manifolds is supersymmetric. Note however that this entire argument depends crucially on the fact that $F = 0$. 

As it turns out, flux is an intrinsic part of most realistic backgrounds, for instance those generated by BPS M-branes. Being charged gravitating objects, these branes modify any space into which they are placed, warping the geometry and giving rise to a field strength flux. The resulting 'back-reacted' manifold is no longer Ricci-flat nor does it admit covariantly constant spinors, but since the brane is BPS, the background remains supersymmetric. 

The work reviewed here is part of a scheme to characterize the geometries of BPS M-branes and arrive at an exhaustive classification of these flux-filled, yet still supersymmetric, backgrounds. 

\section{The Geometry of BPS M-branes}

Before we discuss M-branes wrapping more complicated supersymmetric cycles, we look at a planar M5-brane\footnote{We are mentioning only M5-branes explicitly but of course a parallel analysis can - and in fact has been - carried out for M2-branes as well.} in order to build an intuition for  the general features exhibited by supergravity solutions. Consider an M5-brane placed in Minkowski space.  Spacetime, after being curved by the brane, is described by the metric $ds^2 = H^{-1/3} \eta_{\mu \nu} dX^{\mu} dX^{\nu} + H^{2/3} \delta_{\alpha \beta} dX^{\alpha} dX^{\beta}$ and field strength $F_{\alpha \beta \gamma \delta} = \frac{1}{2}  {\epsilon}_{\alpha \beta \gamma \delta \rho} \; \partial_{\rho} H$. Poincare invariance on the world-volume implies that H be independent of coordinates $X^{\mu}$ tangent to the brane. Rotational invariance in the transverse directions $X^{\alpha}$ says that $H$ can depend only on the radial coordinate $r = \sqrt{X_{\alpha} X^{\alpha}}$ and the metric is diagonal in this subspace. The conditions $dF =0$ and $d*F=0$ fix H to be a harmonic function in the space transverse to the brane, i.e, $H = 1 + \frac{c}{r^3}$. 

More complicated BPS configurations can be generated by wrapping M-branes\footnote{An M2-brane along the $012$ directions preserves the 16 supersymmetries which survive
${\Gamma}_{012} \chi = \chi$. The metric 
and field strength are specified by
$ds^2 = H^{-2/3} \eta_{\mu \nu} dX^{\mu} dX^{\nu} +
H^{1/3} \delta_{\alpha \beta} dX^{\alpha} dX^{\beta}$ 
and $F_{0 1 2 \alpha} =
\frac{ \partial_{\alpha} H}{2 H^2} $ where $H = 1 + \frac{a}{r^6}$, $\mu = 0,1,2$ spans the worldvolume and 
$r$ is the radial coordinate in the transverse space spanned by $\alpha = 3, \dots 10$.} on supersymmetric cycles in a compactification manifold ${\cal M}$. The very presence of the brane deforms space-time such that ${\cal M}$ no longer has special holonomy. Regardless of the particular cycle it wraps, there are some univeral features exhibited by the geometry of a BPS brane wrapped on a supersymmetric cycle $\Sigma$ embedded in a manifold ${\cal M}$.  Space-time splits naturally into three subspaces; unwrapped directions $X^{\mu}$ transverse to ${\cal M}$ but tangent to the brane, ${\cal M}$ itself, spanned by $dX^I$ and the $X^{\alpha}$ directions transverse to both brane and manifold. The flat worldvolume directions exhibit Poincare invariance so nothing can depend on $X^{\mu}$. Along directions $X_{\alpha}$, the brane appears point-like and the configuration is rotationally symmetric. This leads to a diagonal metric in this subspace and further dictates that the functional dependence of any physical quantity only involve the radial coordinate $r$.  Fayyazuddin and Smith incorporated these isometries\footnote{The metric in ${\cal M}$  is left completely general, since we have made no assumptions about either this manifold or the supersymmetric cycle yet.} into the metric ansatz \cite{fs}
\begin{equation}
ds^2 = H_1^2  \eta_{\mu \nu }dX^{\mu} dX^{\nu} + G_{I J } dX^{I}
dX^{J} + H_2^2 \delta_{\alpha \beta} dX^{\alpha} dX^{\beta}
\end{equation}
They then began their search for a supersymmetric supergravity solution by imposing $\delta_{\chi} \Psi = 0$. This lead to a set of equations which can be solved by expressing the field strength and the metric in terms of a single function H. If, in addition $dF=0$ and $d \star F =0$, we are guaranteed that Einstein's equations are satisfied. While the former condition is trivially satisfied, the latter leads to a non-linear differential equation\footnote{H can hence be thought of as the appropriate generalisation (for a wrapped brane case) of the harmonic function in the solution for a planar brane.)} for H. 

Strictly speaking, a supergravity solution is not found until this equation is solved; though possible in theory, this proves very difficult in practise. However, the process we have just described yields an unexpected boon. In addition to relating the metric to the field strength, $\delta_{\chi} \Psi = 0$ also imposes constraints on certain differential forms in ${\cal M}$. It is well known that special holonomy manifolds can be defined through conditions on the differential forms they admit. Proceeding in analogy we can begin to characterize ${\cal M}$ using the constraints obtained by the above analysis.

\section{How BPS M-branes deform Calabi-Yaus}

In the following, we will restrict ourselves to M-branes wrapped on supersymmetric cycles in Calabi-Yau manifolds. A Calabi-Yau n-fold is, by definition, a Ricci-flat K{\"a}hler manifold, with SU(n) holonomy. Equivalently, a Calabi-Yau is defined through the conditions $dJ = d \Omega = 0$, where $J$ is the  K{\" a}hler form and $\Omega$ the unique $(n,0)$ form on the manifold. Even when explicit metrics are not known, as is the case for most Calabi-Yaus, these conditions on the differential forms provide a wealth of information about the geometry of the manifold. The supersymmetric cycles of a Calabi-Yau are even-dimensional holomorphic cycles, calibrated by the appropriate power of $J$, and a Special Lagrangian $n$-cycle calibrated by $\Omega$.  Branes wrapped on these cycles saturate the BPS bound so their charges are fixed by their masses and the calibrating forms are simply the respective volume forms \cite{calibrations}. 

Into spacetime of the form ${\bf R}^{(1,3)} \times$ CY2 $\times {\bf R}^3$, we introduce an M5-brane with worldvolume ${\bf R}^{(1,3)} \times \Sigma_2$ where $\Sigma_2$ is a holomorphic 2-cycle and consequently, calibrated by J. This M5-brane deforms the four-manifold such that it is no longer Calabi-Yau  \cite{holomorphic} but instead satisfies the constraint $\partial_{\cal M} [H^{1/3} * J] = 0$.

If spacetime looked like ${\bf R}^{(1,3)} \times$CY3$\times {\bf R}$, the presence of an M5-brane with worldvolume ${\bf R}^{(1,3)} \times \Sigma_2$ would modify the geometry and the six-manifold would be subject to the constraint $\partial_{\cal M} [H^{- 1/3} * J] = 0$. An M5-brane wrapping a holomorphic 4-cycle (calibrated by $J \wedge J$) can only be non-trivially embedded into a CY 3-fold. The brane deforms this six dimensional space in such a way that the Calabi-Yau condition $dJ = 0$ is replaced by $\partial_{\cal M} [H^{1/3} * J \wedge J] = 0$. When membranes wrap holomorphic two-cycles in CY $n$-folds, their back-reactions distort the 2$n$-dimensional manifolds such that $d _{\cal M} [H^{(n-4)/3} * J] = 0$.

Recall that Calabi-Yau manifolds also admit another kind of supersymmetric cycle, the Special Lagrangian (SpelL). A SpelL $n$-cycle ${\cal L}$, calibrated by $\Omega e^{i \theta}$ is an $n$-dimensional real submanifold of $C^n$ on which $J|_{\cal L} = 0$ and $\Omega e^{i \theta}|_{\cal L} = Vol{(\cal  L})$. 
Even though a phase can be incorporated in general, for simplicity, we will consider SpelL cycles calibrated by $Re \Omega$ [or $Im \Omega$]. 
Since a SpelL is a real manifold, there is no reason to assume that a brane wrapping it will deform space in such a way that a complex structure survives on the backreacted manifold ${\cal M}$. In fact, the Fayyazuddin-Smith analysis of these geometries \cite{SpelLs} shows us that only an {\it almost} complex structure $J$ survives. For a M5-brane wrapping the SpelL 3-cycle\footnote{Since a SpelL 2-cycle in a CY 2-fold is merely a holomorphic 2-cycle in a redefined complex structure, we will mention only SpelL 3-cycles here} calibrated by $Re \; \Omega$, we find that the Calabi-Yau 3-fold is deformed into a manifold on which $\partial_{\cal M}  * \; [Re \; \Omega] = \partial_{\cal M} \; [Im \; \Omega] = 0$. The requirement of   supersymmetry $\partial \Psi = 0$ leads to other constraints as well on $\Omega$ and $J$. Once again, all physical quantities can be expressed in terms of a single function H, which is subject to a non-linear differential equation. 

\section{Conclusions}

In backgrounds without flux, bosonic supersymmetric solutions of supergravity are given simply by metrics with special holonomy. The question is, how do we generalise this classification when flux is non-vanishing? 
As a case in point, we studied the background generated by a M-brane wrapping a holomorphic cycle in a Calabi-Yau manifold. Even though we were able to specify the metric and field strength only modulo solution of a non-linear differential equation, we found certain defining equations for the manifold ${\cal M}$, into which the Calabi-Yau was deformed. We then proceeded to study M-branes wrapping other supersymmetric cycles, aiming not to find explicit metrics for a handful of examples but intead to characterise a back-reacted manifold through constraints on its differential forms. Our approach differs from the usual in that most of the work done in this field focuses on finding explicit supergravity solutions in some approximation; this solutions will at best tell us about the local geometry near the cycle. In  contrast, the statements we make are global and can be used to classify the back-reacted manifold.  For all M-branes wrapping supersymmetric cycles in Calabi-Yau manifolds\footnote{The only exception is the M5-brane wrapping a holomorphic 4-cycle in a CY 4-fold \cite{bt}. It has been known for a while that this configuration stands apart from the crowd in a number of ways, a  key reason behind this apparent discord being that it is the only M-brane geometry encountered so far which does not satisfy $F \wedge F = 0$}  we obtain a constraint on the {\bf dual} of (the appropriately rescaled) calibration on ${\cal M}$.  Such constraints, we hope, will lead to a concise and exhaustive classification of supersymmetric flux backgrounds.\\

\noindent
{\bf Acknowledgments:} I am grateful to the organizers for the stimulating atmosphere they created in Islamabad this April. Faheem Hussain, in particular, went gone out of his way to make the conference a success. I  would also like to thank Ansar Fayyazuddin for several vary enjoyable collaborations on which this review is based.

\end{document}